\documentclass[12pt]{article}
\usepackage{epsfig}
\begin{document}
\centerline{\bf Variational Approach to Hard Sphere}
\centerline{\bf Segregation Under Gravity}
\vskip 1.0cm 
\centerline{ Joseph A. Both and Daniel C. Hong}
\centerline{ Physics, Lewis Laboratory, 
Lehigh University, Bethlehem, Pennsylvania 18015}
\begin{abstract}

It is demonstrated that the minimization of the free energy functional
for hard spheres and hard disks yields the result that excited granular 
materials under gravity segregate not only in the widely known 
``Brazil nut'' fashion, i.e. with the larger particles rising to the top, 
but also in reverse ``Brazil nut'' fashion.  Specifically, the local
density approximation is used to investigate the crossover between 
the two types of segregation occurring in the liquid state, and the
results are found to agree qualitatively with previously published results 
of simulation and of a simple model based on condensation.

\end{abstract}



Segregation of hard sphere mixtures has a long history, starting perhaps from
 the classic papers by Wood and Jacobson [1], followed by Lebowitz, 
Rowlinson, and Widom  [2,3], extending to the most recent works by various 
groups [4].  It is still controversial whether or not hard sphere mixtures 
do phase segregate in the absence of gravity [4].  However, in the presence 
of gravity, the situation is rather clear cut.  
Rosato et al. [5] advanced an explanation of the segregation 
in vertically shaken mixtures of granular materials
by appealing to
 geometrical reorganization: during shaking,
voids opening beneath larger particles are filled more readily 
by smaller particles.  This may be contrasted
with the apparent 
``buoyancy'' of larger particles [6], or
 convection driven 
segregation [7].  But recent molecular dynamics simulations of 
weakly dissipative
 hard sphere and hard disk granular systems under gravity that are in global
 thermal equilibrium with a heat reservoir have shown that a reverse 
segregation phenomenon occurs as well [7].
These results were interpreted in light of a recent proposal that the hard 
spheres undergo condensation transition under gravity [8], which was
 subsequently tested by Molecular Dynamics simulations [10]. 

 The density profile in the condensed regime is fairly uniform at the 
level of Enskog approximation [8], or displays oscillatory structure in
 the weighted density functional approximation [9, 11], which is 
consistent with the experimental observation of the formation of crystalline
 structure with fairly uniform density near the bottom of the shaken granular
 materials [18].   
It was argued that the type of segregation
 (Brazil nut [BN] or reverse Brazil nut [RBN]) results from a
 competition between the system's tendency to reorganize geometrically in 
the ways described by Rosato et al. [5] and the tendency of hard spheres 
under gravity to condense at low enough kinetic temperature [8-10]. 
It was suggested as a 
qualitative model [7] that in a binary hard sphere mixture, the particles 
belonging to the different species would be characterized by 
different condensation temperatures, so that
 during quenching from high to low temperature, the species with highest  
 would tend to condense at the bottom of the sample first, leading ultimately
 to vertical segregation if this tendency were realized. This condensation 
driven tendency toward segregation, however, would be either augmented or 
opposed by the essentially geometric mechanisms identified by Rosato et al.,
 which favor BN segregation only, depending respectively upon whether the 
smaller particle has higher or lower condensation temperature   than the 
larger particle.  
The purpose of this Letter is to present for the first time
a theory based on the variational principle
and directly derive the
phase boundary of
the binary mixtures.

{\bf Theory} : 
We calculate in the local density approximation (LDA).
Specifically, with $\psi_{id}$ being
the Helmholtz free energy per particle for
the ideal gas in the absence of gravity, 
and $\psi_{exc}$ being that of the excess component 
due to particle interactions, the binary 
mixture free energy 
 per area functional is given as a function of the densities $\rho_i({\bf r})$:

\begin{eqnarray}
\bar F[\rho_1(z),\rho_2(z)]& = &
 \int_0^\infty dz\,\rho(z) \psi_{id}(\rho_1,\rho_2) + \nonumber \\
& & \int_0^\infty dz\, \rho(z) \psi_{exc}(\rho_1,\rho_2) + \nonumber \\
& & m_1 g\int_0^\infty dz\,\rho_1 z +  m_2 g\int_0^\infty dz\,\rho_2 z,
\label{F_LDAbin}
\end{eqnarray} 
where 1 and 2 are particle indexes, and the total
density, $\rho$, is the sum of the two: $\rho = \rho_1 + \rho_2$.
This gives the free energy of a columnar sample of the system whose 
transverse cross section has one unit of area.  The plane $z=0$ 
is the bottom of the container, and 
for the  problem to be strictly one dimensional, the size of the container
in the transverse direction must be infinite.  Minimizing $\bar F$ 
under the global constraints that the number of particles
of each species is conserved:
\begin{equation}
N_i = \int_V d{\bf{r}}\rho_i({\bf {r}}),\,\,\,\,\,\,\, i=1,2
\label{PARTNUM}
\end{equation}
yields the desired density profiles.

The forms of $\psi_{id}$ are given in [12]
and it is standard [12, 13] that
if 
$Z \equiv \frac{P}{\rho T}$, then
\begin{equation}
\frac{\psi_{exc}}{T} = \int_0^\rho\left(Z(\rho^\prime)
-1\right)\frac{d\rho^\prime}
{\rho^\prime}.
\label{PSI_EXC}
\end{equation}
For the form of $Z$, we use a generalized Carnahan-Starling equation of 
state, the so called Mansoori, 
Carnahan, Starling, Leland approximation for a finite number of hard 
sphere species [14].
Defining 
$\xi_{\alpha} = \frac{\pi}{6} \sum_{i=1}^{n} \rho_i D_i^\alpha\,\,\,\,\,\,\,
(\alpha = 0,...,3)$,
with $\rho_i = N_i/V$, and $n$ being the number of species in the
mixture,  $Z$ is then given by
\begin{equation}
Z = 
\frac{6}{\pi\rho}\left[\frac{\xi_0}{1-\xi_3} + \frac{3\xi_1\xi_2}{(1-\xi_3)^2}
+ \frac{3\xi_2^3}{(1-\xi_3)^2} - \frac{\xi_3\xi_2^3}{(1-\xi_3)^3}\right],
\label{Z}
\end{equation}
where $\rho = \sum_{i=1}^{n} \rho_i$.
To perform the integration specified in Eq. (\ref{PSI_EXC}) for
$n=2$, we recognize
that this is an expression for the excess free energy for a 
homogeneous
mixture in which $\rho_1$ and $\rho_2$ are {\it fixed} fractions of $\rho$, 
i.e. $\rho_1/\rho$ and $\rho_2/\rho$ are both constants.
Then one readily sees then that the quantities $\xi_i$ may be written as 
constant multiples of
$\rho$ only, viz. 
$\xi_{\alpha} = C_{\alpha} \rho,$
which defines the constant $C_{\alpha}$ in terms of $\rho_1,\rho_2$ and $D_1$
 and $D_2$.

Performing the integration specified in Eq. (\ref{PSI_EXC})
with the substitution of these relations, we find the desired form for 
$\psi_{exc}$: 
\begin{equation}
\frac{\psi_{exc}}{T} = \left(\frac{\xi_2^3}{\xi_0\xi_3^2} - 1 \right)
\ln(1-\xi_3)+\frac{3\xi_1\xi_2}{\xi_0(1-\xi_3)}
+\frac{\xi_2^3}{\xi_0\xi_3(1-\xi_3)^2},
\label{PSI_EXC2}
\end{equation}
(see refs. [15] and [16]).
This done, the functional $\bar F[\rho_1(z),\rho_2(z)]$ defined in Eq. 
(\ref{F_LDAbin}) is then fully specified. 
Its minimization is accomplished by solving for $\rho_1(z)$ and $\rho_2(z)$ 
the 
two equations ($i=1,2$):
\begin{equation}
\frac{\delta \bar F}{\delta \rho_i}=
\frac{d\left[ \rho\psi_{id}\right]}{d\rho_i} + 
\psi_{exc} + \rho
 \frac {d\psi_{exc}}{d\rho_i}
+ m_igz = \lambda_i,
\label{ALGEQNS}
\end{equation}
where the Lagrange multipliers $\lambda_i$ are introduced to constrain the
minimization according to Eq. (\ref{PARTNUM}).  
These equations, 
though only algebraic, are non-linear and highly nontrivial and must 
 therefore be
solved numerically 
via an iterative scheme for $\rho_i(z)$ at several points $z$.  
The two dimensional problem is solved analogously.  We use the equation of
state of hard disks used by Jenkins and Mancini [17]
\begin{equation}
Z = 1 + \frac{K_{11}}{\sigma_1^2} + 8 \frac{K_{11}}{\sigma_{12}^2}
+\frac{K_{22}}{\sigma_2^2},
\label{JMEOS}
\end{equation}
where $\sigma_{i} = D_{i}/2$ and $\sigma_{ij} = \sigma_i+\sigma_j$,
and where
\begin{equation}
K_{ij} = \frac{\pi}{8}\rho_i \rho_j \sigma_{ij}^4 g_{ij}
\label{KIJ}.
\end{equation}
The functions $g_{ij}$ are the pair correlations at contact.  Following Jenkins
and Mancini [17], we use
\begin{eqnarray}
g_{11}& =& \frac{1}{1-\eta} + \frac{9}{16} \frac{\eta_1 + \eta_2 R}{(1-\eta)^2}
\nonumber \\ 
g_{22}&=& \frac{1}{1-\eta} + \frac{9}{16} \frac{\eta_1 + \eta_2 R}{R(1-\eta)^2}
 \\ 
g_{12}&=& \frac{1}{1-\eta} + \frac{9}{8} \frac{\eta_1 + \eta_2 R}
{(1+R)(1-\eta)^2} \nonumber
\label{GIJ}
\end{eqnarray}
where $\eta_i = \frac{\pi}{4} D_i^2 \rho_i$, the area fraction of species $i$,
$\eta = \eta_1 + \eta_2$, and $R = D_1/D_2$.  Eq. (\ref{PSI_EXC}) 
is integrated,
again using the fact that the fractions $\rho_i/\rho$ are constant 
 and then equations analogous to Eq. 
(\ref{ALGEQNS}) are then solved iteratively for $\rho_i(z)$, where the
derivatives of $\psi_{exc}$ are evaluated numerically.

\begin{figure}
\begin{center}
\epsfig{file=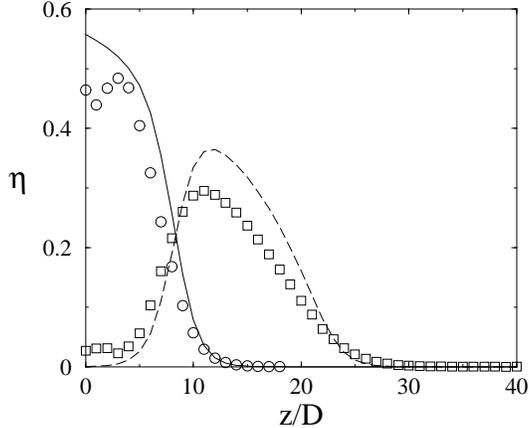,height=6cm,angle=0.,clip=}
\end{center}
\caption{A comparison of density profiles calculated from the LDA (lines)
and from molecular dynamics simulation (symbols).  The dashed line and
squares designate species 1, and the solid line and circles designate 
species 2; $m_2/m_1 = 2$; $D_2/D_1 = 1$; $\mu_1\approx\mu_2\approx 8$.}
\label{COMPFIG}
\end{figure}
{\bf Results and Discussion}: 
The control parameters that directly enter this formulation of the problem 
are $g$, $T$, $m_i$, $D_i$, and $\lambda_i$.  In practice, for any
choice of $g$, $T$, $m_i$, $D_i$, and $\mu_i$, the last of which
control the average particle density,  the parameters 
$\lambda_i$ are tuned so that the integrated density profiles 
yield desired layer numbers $\mu_i$ according to 
$\mu_i = D_i^{d-1}\int dz \rho_i$, where $d$ is the dimensionality of the 
system.  To show that the LDA 
is capable of generating reliable results, we present Fig. \ref{COMPFIG}, which
shows volume fraction profiles $\eta_i(z) = \frac{\pi}{6}D^3\rho_i(z)$ for 
$d=3$, $D_1 = D_2 = 0.001$ m, $m_1 = 1.047\times10^{-6}$ kg, $m_2 = 2m_1$,
$g = 9.8$ m/s$^2$ generated by the LDA (lines) and generated by MD simulation 
(symbols) used in ref. [7].
  The layer numbers $\mu_1$ and $\mu_2$ are both nominally $8$, 
but the integrals of the LDA curves disagree slightly with those of the MD 
curves as a result of a systematic binning error in the MD 
results.  The temperatures of the two systems differ; $T_{LDA}/
m_2gD_2\mu_2 = 0.0731$ and  $T_{MD}/m_2gD_2\mu_2 = 0.1401$.  The LDA system 
is not
at all dissipative, so it is reasonable that the different systems give
similar profiles only at {\it different} temperatures, 
$T_{MD}$ being necessarily the higher one.  We are, however, not able
to exactly pin point the factor
two difference in temperatures.  The temperature difference 
notwithstanding, the resulting profiles
are very similar, and indeed exhibit non-BN segregation.  Note that 
in this case, 
the segregation mechanism cannot be 
geometric, because the particles are of equal size but with different mass. 

In all subsequent work reported here, we choose $g = 10$ m/s$^2$, 
$D_1 = 0.001$ m, $m_1 = 10^{-6}$ kg, and $\mu_i = 10$ to within less 
than one tenth percent.  To investigate the dependence of segregation on
mass and diameter ratios, we vary $D_2$ and $m_2$ such that $D_2\ge D_1$
and $m_2\ge m_1$, 
find $\eta_i(z)$ at low 
temperature, and quantify the segregation as a function of $D_2/D_1$, 
$m_2/m_1$, and $T$.  Our choice to quantify the segregation is simply the
ratio of the centers of mass of the two species: ${\langle z_1\rangle }/{\langle z_2\rangle }$.
Specifically, ${\langle z_1\rangle }/{\langle z_2\rangle } < 1$ indicates BN 
segregation, ${\langle z_1\rangle }/{\langle z_2\rangle } = 1$ indicates
crossover, and ${\langle z_1\rangle }/{\langle z_2\rangle } > 1$ 
indicates RBN segregation. 
To reduce the number of control parameters and to try to make 
some comparison
with the condensation theory [7]
which originally motivated this work, we choose to
explore the lower temperature regimes at a fixed {\it reduced} temperature.  Because the LDA cannot generate information about the 
solid phase, we are careful to choose temperatures  at which total 
volume fractions, $\eta(z)=\eta_1(z) + \eta_2(z)$ have their maximum 
somewhere on the range
$0.50$ to $0.65$ when $d=3$, that is, near the density of the simple cubic packed single 
species solid ($\eta \approx 0.52$), but not as dense as 
the hexagonally packed single species solid ($\eta \approx 0.74$).  When $d=2$,
 the maximum of $\eta(z)$ is kept on the range $0.80$ to $0.85$ (also between
the square and triangular packing single species area fractions, approximately 
$0.79$ and $0.91$).
Thus, our method explores fluid systems of moderate to high densities near the
bottom of the sample.    We have
found that choosing a reduced temperature $\tilde T \equiv T/(m_2gD_2\mu_2) =
 0.0375$ ($d=3$) and $\tilde T \equiv T/(m_2gD_2\mu_2) = 0.0300$ ($d=2$) 
keeps the maximum of $\eta(z)$ on the prescribed ranges over the spectrum 
of diameter and mass ratios we have used.

\begin{figure}
\begin{center}
\epsfig{file=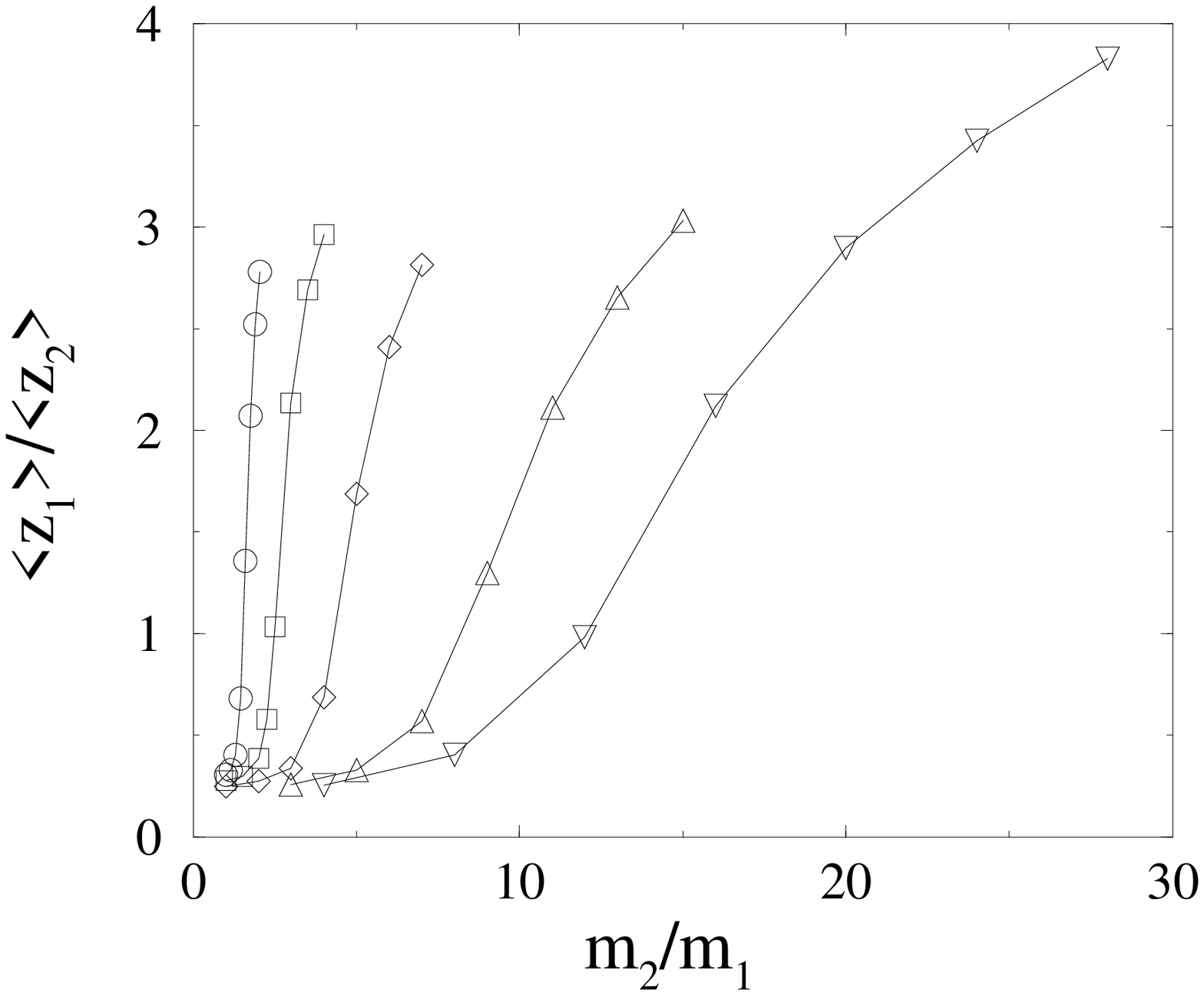,height=6cm,angle=0.,clip=}
\epsfig{file=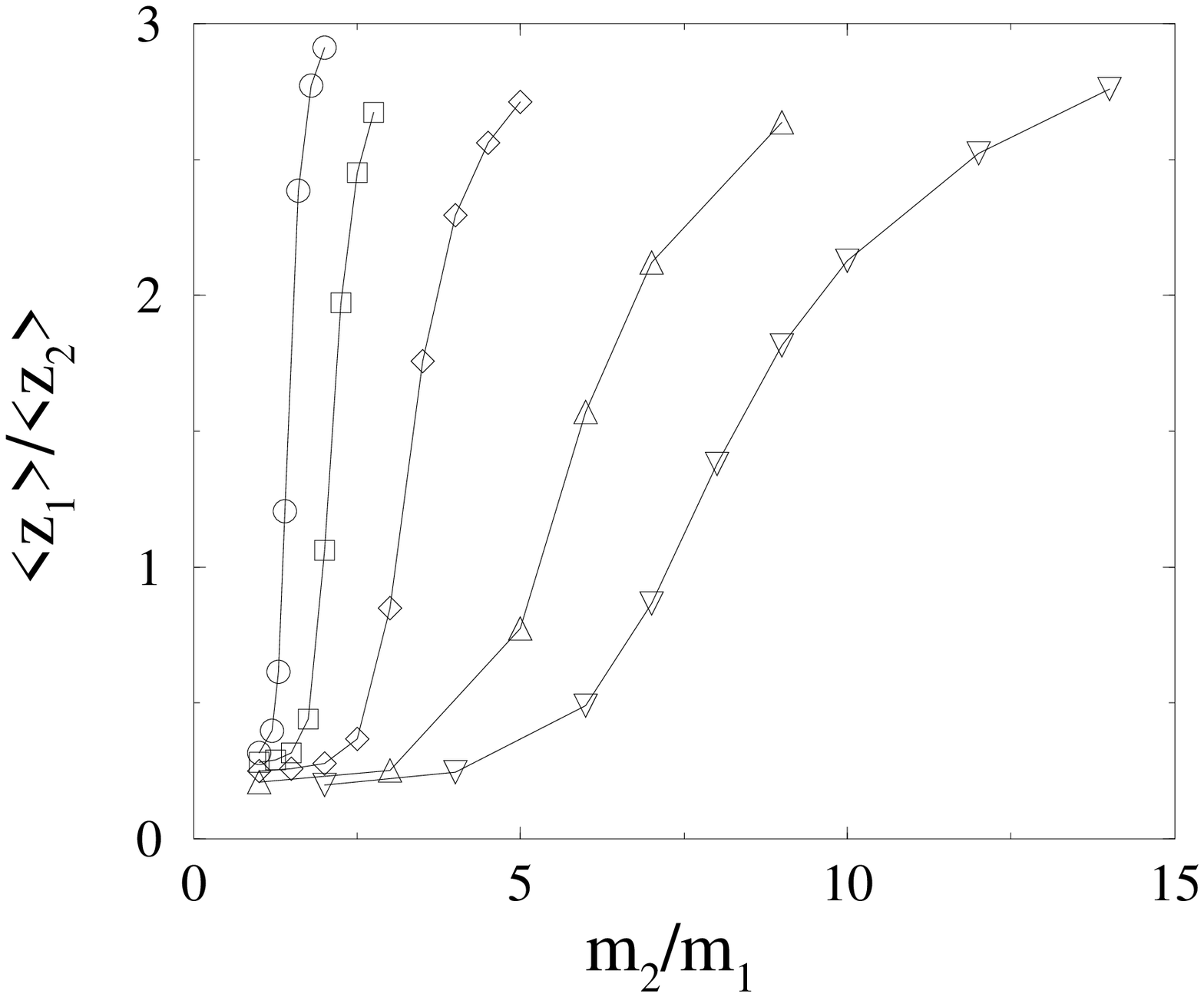,height=6cm,angle=0.,clip=}
\end{center}
\caption{Segregation parameter ${\langle z_1\rangle }/{\langle z_2\rangle }$ as a function of mass
ratio $m_2/m_1$ for hard spheres (left) and hard disks (right).  In both
graphs $\mu_1=\mu_2=10$;
$D_2/D_1 = 1.2$ (circle), 1.5 (square), 2.0 (diamond), 3.0 (up triangle), 
4.0 (down triangle).  $\tilde T_{3d} = 0.0375$; $\tilde T_{2d}=0.0300$. }
\label{SEGPARAMFIG}
\end{figure}

Fig. \ref{SEGPARAMFIG} summarizes the
results of our calculations for diameter ratios 
\begin{figure}
\begin{center}
\epsfig{file=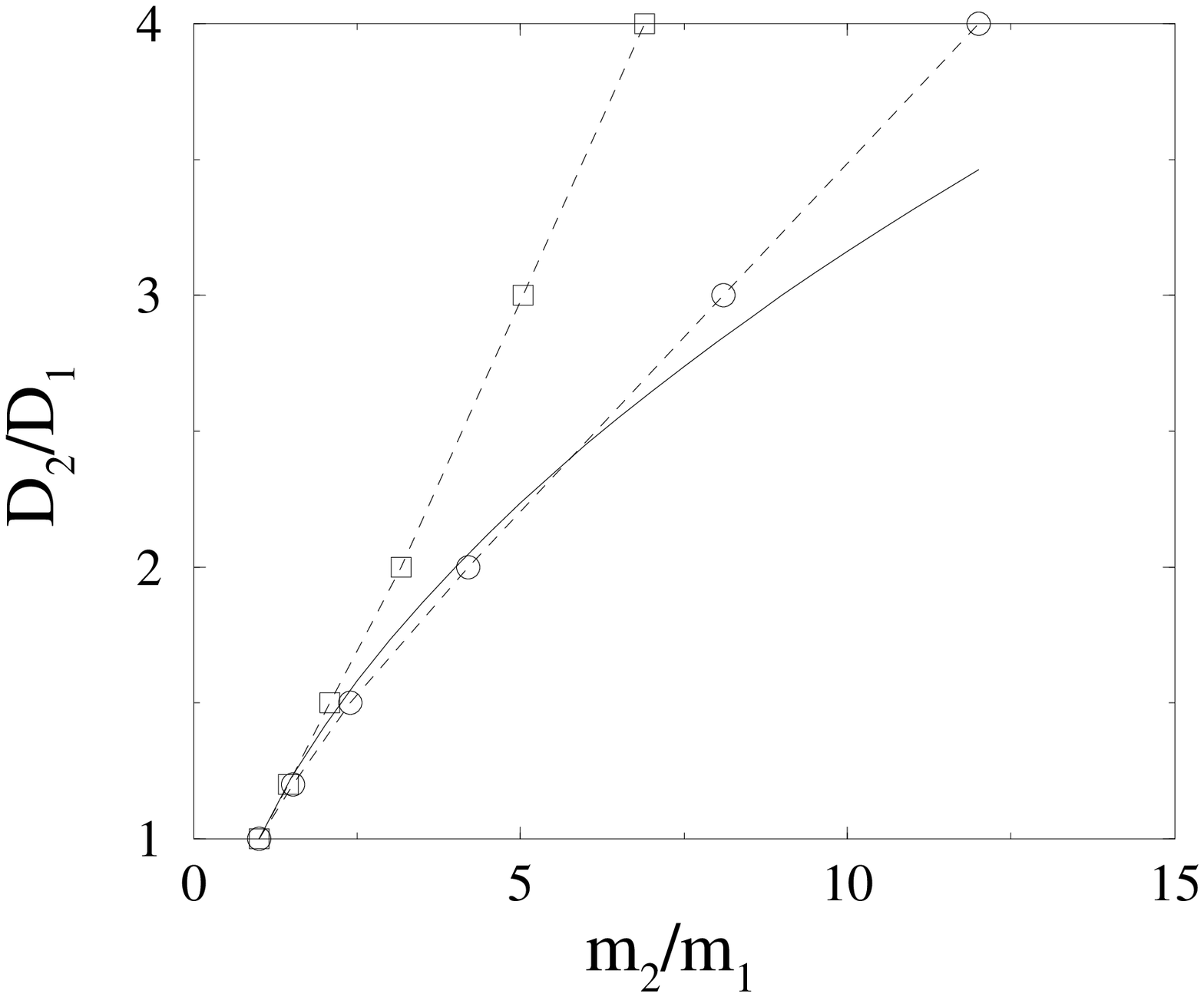,height=6cm,angle=0.,clip=}
\epsfig{file=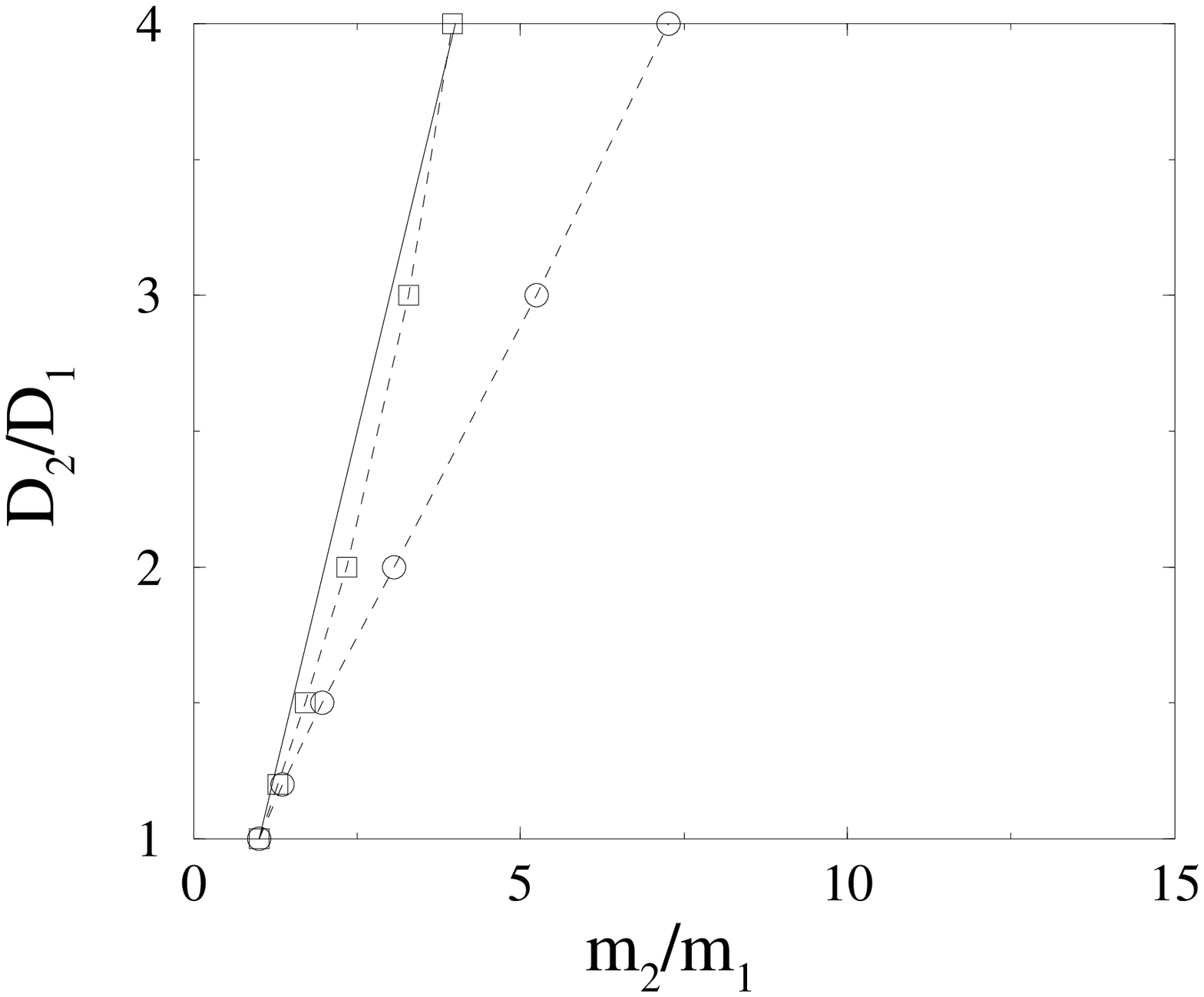,height=6cm,angle=0.,clip=}
\end{center}
\caption{Curves through parameter space along which the segregation 
parameter ${\langle z_1\rangle }/{\langle z_2\rangle }=1$, 
indicating the crossover from BN (left) to 
RBN (right) segregation, for $d=3$ (upper) and for $d=2$.  
The dashed curves marked by circles are from linear 
interpolation of the data shown in Fig. {\ref{SEGPARAMFIG}},
 and are for the low temperatures
given in the text.
The dashed curves marked by squares are for higher reduced 
temperature ($\tilde T = 0.1500$ 
in both cases).} 
\label{PHASEFIG}
\end{figure}
$D_2/D_1 = 1.2,\,1.5,\,2.0,\,3.0,\,4.0$, and a range of mass ratios for both
$d=2$ and $3$.    If from these data
we interpolate to find those values of $m_2/m_1$ at which crossover occurs 
($\langle z_1\rangle /\langle z_2\rangle  = 1$) for given values of $D_2/D_1$, we generate the data 
shown
in Fig. \ref{PHASEFIG}.  Each graph displays three curves.  The dashed curves
with symbols are the crossover curves calculated with the LDA at the 
low temperatures
cited above and also at $\tilde T = 0.1500$ for both $d=2$ and $d=3$, so that
the temperature dependence may be illustrated.
To the left of a crossover curve, 
the segregation is of the BN variety; to the
right, it is RBN.  Thus, the LDA reveals that as reduced temperature increases,
an increasingly larger region of parameter space gives rise to RBN segregation.
Indeed, if we assume that in the high temperature limit the density of
each species is proportional to $\exp(-m_i gz/T)$, then for $\mu_1$ = $\mu_2$,
one may calculate for $d=2$ and $d=3$ that the ratio of centers of mass is
given by
\begin{equation}
\frac{\langle z_1\rangle }{\langle z_2\rangle }=\frac{m_2}{m_1},
\label{HITEMPCROSS}
\end{equation} 
which indicates the crossover curve is the vertical line $\frac{m_2}{m_1} = 1$,
a result consistent with the trend in the LDA results presented in 
Fig. \ref{PHASEFIG}.

Also appearing on the both graphs in Fig. \ref{PHASEFIG} as solid lines are 
crossover curves predicted from
the simple condensation argument.  Briefly, these are obtained by assuming
the ratio of condensation temperatures in the single species theory
$$T_{c2}/T_{c1} = m_2gD_2\mu_2/m_1gD_1\mu_1$$ measures the tendency of species
 2 to 
segregate to the bottom at $T<T_{c2}$ due to the condensation mechanism [7].  

If we choose to equate this with Rosato et al.'s [5] control parameter
for the reorganization,
 $(D_2/D_1)^d$, we can obtain the
crossover curve (The solid lines in Fig. 3).
 We see fair agreement with our 
theoretical prediction at low reduced temperature 
for $d=3$, but significant disagreement
appears at $m_2/m_1 \approx 7$; the crossover curve seems practically linear, 
in disagreement with the scaling suggested by the condensation argument.
For $d=2$, the low temperature LDA boundary is linear, in agreement with the 
scaling suggested by the condensation argument, but that argument of course 
cannot predict the slope.  Moreover, the LDA computed boundaries 
are necessarily temperature dependent, but the  
condensation theory [7]
does not predict a temperature dependence.  Thus, although
the condensation theory has been useful in directing us to look for certain
dependencies and, indeed, for the phenomenon of RBN segregation itself, its
utility, except as a heuristic, is somewhat circumscribed.  However, the 
equilibrium approach of 
minimizing Helmholtz free energy seems a promising theoretical alternative,
as it is capable of generating solutions in good agreement with those obtained
from MD simulation, as in Fig. \ref{COMPFIG}.


\vskip 1.0 true cm

\noindent {\bf References}

\noindent [1] W.W. Wood and J. D. Jacobson, J. Chem. Phys. 
{\bf 27}, 1207 (1957).
 
\noindent [2] J. Lebowitz and J.S Rowlinson, J. Chem. Phys. {\bf 41}, 
133 (1964).

\noindent [3] B. Widom and J. S. Rowlinson, 
J. Chem. Phys. {\bf 52}, 1670 (1970).

\noindent [4] M. Dijikstra, R. von Roji, and R. Evans. Phys. Rev. E. 
{\bf 51}, 5744 (1999).

\noindent [5] A. Rosato, et al., Phys. Rev. Lett. {\bf 58}, 1038 (1987).

\noindent [6] T. Shinbrot and F. Muzzio, Phys. Rev. Lett. {\bf 81}, 
4365 (1998).

\noindent [7] D. C. Hong, P. V. Quinn and S. Luding, 
Phys. Rev. Lett. {\bf 86}, 3423 (2001).

\noindent [8] D. C. Hong, Physica A {\bf 271}, 192 (1999).

\noindent [9] J. A. Both and D. C. Hong, Phys. Rev. E {\bf 64}, 061105 (2001).

\noindent [10] P. V. Quinn and D. C. Hong, Phys. Rev. E {\bf 62}, 8295 (2000).

\noindent [11] P. Tarazona, Mol. Phys {\bf 52}, 81 (1984).

\noindent [12] P. M. Morse, {\it Thermal Physics}, Addison-Wesley Publishing 
Co., Redwood City, CA, 1969.

\noindent [13] J. P. Hansen and I. R. McDonald, {\it Theory of Simple
Liquids}, Academic Press, New York, 1976.

\noindent [14] G. A. Mansoori, et al., J. Chem. Phys. {\bf 54}, 1523 (1971).

\noindent [15] J. J. Salacuse, G. Stell, J. Chem. Phys {\bf 77}, 3714 (1982).

\noindent [16] A. Meroni, A. Pimpinelli, L. Reatto, J. Chem. Phys {\bf 87},
3644 (1987).

\noindent [17] J. T. Jenkins and F. Mancini, J. Appl. Mech. {\bf 54}, 29,
(1987).

\noindent [18] T. Biben, R. Ohnesorge and H. Lowen, Europhys. Lett. {\bf 71}, 615 (1993).
\end{document}